\documentstyle[setspace,11pt,amsmath,amssymb,rsfs]{article}

\textheight 
9.5in
\textwidth 
6.5in 
\oddsidemargin 
-0.25in
\topmargin 
-0.5in

\begin{document}
\onehalfspacing
\begin{titlepage}

\flushright{\today}

\vspace{1in}

\begin{center}

{{\textbf{Dynamical Gauged S-duality Spontaneous Breakdown
}}}

\vspace{1in}

\normalsize

{Eiji {{Konishi}}\footnote{konishi.eiji@s04.mbox.media.kyoto-u.ac.jp}}

\normalsize
\vspace{.5in}

 {\em Department of Physics, School of Science, Kyoto University\\ Sakyo-ku, Kyoto 606-8502, Japan}

\end{center}

\vspace{1in}

\baselineskip=24pt
\begin{abstract}
Inspired by Yoneya's recent work on $D$-brane field theory, we present the constructive definition of this theory as a new dual model based on the quantization of non-perturbative string gauged $S$-dualities and its spontaneous breakdown. Our model is determined by one BRST quantization condition. Here, we provide some testable conjectures, for example, that the partition function of bosonic part of IIB matrix model and string partition functions of various interactions including gravity are realized by the physical state and its Goldstone modes. 
\end{abstract}

\end{titlepage}
In this letter, we consider a model which incorporates the dynamical gauged $S$-duality of type IIA/$M$ theories.\cite{KM} As the dynamical objects, we select the $D$-particle fields proposed recently by Yoneya.\cite{Y1}

First, we explain the idea of {{gauged}} $S$-duality. In type IIB theory, for instance, $S$-duality acts on the string coupling constant $g_s=\chi+\sqrt{-1}e^{-\phi}$ for the axion-dilaton $(\chi$-$\phi)$. Usually $S$-duality is considered as a non-linear {{global}} symmetry on the Poincar\'e upper half plane $\mathfrak{h}$. In contrast, we take the gauge transformation on each vacuum on $\mathfrak{h}$ to be independent of the others. Using this idea and Yoneya's $D$-particle field theory, we can consider the dynamics of duality on hyperbolic curves parameterized by the coupling constant $g_s$.

Our formulation of $D$-particle fields resembles the dual models (Regge pole theory) studied in the 1960s and 70s: in Regge pole theory, the scattering amplitudes between two hadrons are plotted as bound states on the Cauchy plane or Riemann surface. In Regge theory, a curve plays a mathematical role which fully represents the essence of the physics. There are only two parameters that control the theory: the slope parameter of the Regge trajectory, $\alpha^\prime$, and the coupling constant $g_s$.\cite{R}

Yoneya, Scherk and Schwarz have given a dual model description of the duality in electro-gravidynamics by taking the limit $\alpha^\prime \to 0$ and fixing a variable combining the slope parameter and the coupling constant.\cite{Y2}

We formulate a dual-model-like theory for gauged $S$-duality and the spontaneous breakdown of this symmetry. Using this model we can understand the dualities appearing in modern string theory as the result of dynamical gauged $S$-duality spontaneous breakdown (DGSSB).

We now describe the second {{quantized}} $S$-duality wave function. In our physical theory of DGSSB, the dynamical variables are the {{$D$-particle fields}} in typeIIA/$M$ theory. The physical state ${\psi}$ for gauged duality symmetry is the partition function for the ``thermal'' (mass) variable -- the coupling constant $g_s$ -- of the $D$-particle fields in $M$-theory. We take the prototype of $M$-theory to be the Banks, Fischler, Shenker and Susskind (BFSS) matrix model, a one-dimensional reduced model of 10-dimensional supersymmetric Yang-Mills theory. \cite{BFSS}

Our real task is to derive the equation of this physical state.

The gauge symmetry of $S$-duality can be considered as an $SL(2,{\mathbb{R}})$ Yang-Mills theory whose variables are the string coupling constant $g_s$ and deformation time parameters $t_n$ and $\bar{t}_n$, $n=1,2,\cdots$ on a moduli space (curve). Integrablity follows from the fact that the matrix model is integrable and its partition function is a tau function of an integrable system.

We assume that the $D$-particle coupling constant, ${\mathscr{G}}$, is defined by\begin{equation}{\mathscr{G}}\equiv \bigl(\alpha^\prime\bigr)^{-1}.\label{eq:G}\end{equation}

The BRST transformation for the gauged $S$-duality is defined for this coupling constant and three families of {{ghosts}} (two fermionic modes for gauge fixing and a bosonic mode)
\begin{equation}{\mathscr{C}},\ \ \ \ \ \bar{{\mathscr{C}}}\ \ \ \ \ and\ \ \ \ \ \Phi.\end{equation}

The $SL(2,{\mathbb{R}})$ Yang-Mills Lagrangian of time variables $t_n$ in the $R_\xi$ gauge ($\xi\equiv1$) is
\begin{equation}
{\mathscr{L}}\equiv\sum_{n=1}^\infty {\mathscr{L}}_n,\ \ \ \ \ {\mathscr{L}}_n={\rm{tr}}\Biggl[-\frac{1}{4}\Bigl(D_n {\mathscr{A}}^I\Bigr)^2+\partial_n\bar{{\mathscr{C}}}D_n{\mathscr{C}}+\frac{\Phi^2}{2}-i\partial_n\Phi^I{\mathscr{A}}^I\Biggr]
\end{equation} and the corresponding BRST transformation ${\mathbf{U}}$ is\cite{BRST}
\begin{eqnarray}
{\mathbf{U}}\equiv\sum_{n=1}^\infty {\mathbf{U}}_n,\ \ \ \ \ {\mathbf{U}}_n\left\{\begin{array}{c}
{\mathscr{A}}^I\\
{\mathscr{C}}^I\\
\bar{{\mathscr{C}}}^I\\
\Phi
\end{array}\right\}
=\left\{\begin{array}{c}
D_n {\mathscr{C}}^I\\
-\frac{1}{2}({{\mathscr{C}}}\times_G{{\mathscr{C}}})^I\\
 i\Phi^I\\
 0
\end{array}\right\}
\label{eq:trans}
\end{eqnarray}
where \begin{equation}
\partial_n\equiv \partial_{t_n}+\partial_{\bar{t}_n},\ \ \ \ \ D_n\phi\equiv \partial_n\phi+i{\mathscr{G}}{\mathscr{A}}\times_G\phi
\end{equation} and $I$ is the index and $(A\times_GB)^{c}=f_{ab}^{c}A^aB^b$ is the Lie bracket of $G=SL(2,{\mathbb{R}})$.

In the second quantized theory, the gauge potential ${\mathscr{A}}^I$ is expanded in terms of the time variables\footnote{
 These operators, like any operators ${\mathscr{O}}^I$ with an $SL(2,{\mathbb{R}})$ index $I=1,2,3$, satisfy the following $SL(2,{\mathbb{R}})$ commutation relations
\begin{equation}
\bigl[{\mathscr{O}}^I,{\mathscr{O}}^J\bigr]=f^{IJ}_K{\mathscr{O}}^K\end{equation}
with structure constants
\begin{equation}
f_1^{32}=f_2^{31}=f_3^{12}=1,\ \ \ \ \ f_1^{23}=f_2^{13}=f_3^{21}=-1. 
\end{equation}}
\begin{equation}
{\mathscr{A}}=
\sum_{I=1}^3{\varepsilon}^I
\exp\Bigl(\sum_{n=1}^\infty \bigl(t_n{q}^{I}_n +\bar{t}_n\bar{q}^I_n \bigr)\Bigr) 
\end{equation}
for $c$-number $SL(2,{\mathbb{R}})$ matrix ${\varepsilon}^I$ and
 nilpotent operators $q^{I}_n$\begin{equation}\sum_{I=1}^3{\varepsilon}^I\bigl(q_n^I\bigr)^2=0\end{equation}

This expansion is similar to the definition of a time-ordered Green function ($\tau$ function). Our $\tau$ function has infinite number of time variables, $t_n$, which are regarded as time promoting parameters for the conserved charges, ${q}_n$, of the corresponding infinite symmetries. Here the gauge symmetries are Chan-Paton factors on the boundaries of open string fields. This is why we write the operator parts as $\prod_{n}\exp\bigl(t_n{q}^I_n\bigr)$.

The second quantization for the ghosts ${\mathscr{C}}$ and $\bar{{\mathscr{C}}}$ is done in a similar way, conserving the equations of motion.
\begin{equation}
{\mathscr{C}}=\sum_{I=1}^3
{{\varepsilon}}^I
\exp\Bigl(\sum_{n=1}^\infty \bigl(t_n{\theta}_n^I+\bar{t}_n\bar{\theta}_n^I\bigr)\Bigr)\end{equation}
for the same $c$-number $SL(2,{\mathbb{R}})$ matrices
${{{\varepsilon}}}^I$.

It is worth remarking that the $q_n^I$ are fermionic operators while the $\theta_n^I$ are Grassmann numbers.

We define the mass of ${\mathscr{A}}$ by
\begin{equation}
\Delta\Bigl[\vec{t},\vec{\bar{t}}\Bigr]{\mathscr{A}}\biggl|_{\vec{t}=\vec{\bar{t}}=0}=-\Lambda{\mathscr{A}}\biggl|_{\vec{t}=\vec{\bar{t}}=0},\ \ \ \ \ \Delta\Bigl[\vec{t},\vec{\bar{t}}\Bigr]\equiv \sum_{n=1}^\infty \biggl[\frac{\partial^2 }{\partial t_n^2}+\frac{\partial^2}{\partial \bar{t}^2_n}\biggr]\end{equation} The operator $\Delta\Bigl[\vec{t},\vec{\bar{t}}\Bigr]$ is the Laplacian in the time variables $t_n$ on the moduli curve.

If the $SL(2,{\mathbb{R}})$ gauge symmetry of the physical state is not broken, $\Lambda$ must be zero. So, if $\Lambda\neq0$, we regard this violation as the inducer of the spontaneous symmetry breakdown of gauged $S$-duality.

The {{BRST charge}} ${\mathbf{Q}}$ is nilpotent, so this equation is $S$-duality covariant, \begin{equation}[{\mathbf{Q}},\phi]={\mathbf{U}}\phi,\end{equation} for BRST transformations ${\mathbf{U}}$.

The general Kugo-Ojima formula of the BRST charge for our Noether current ${\mathscr{J}}$
\begin{equation}
{\mathscr{J}}\equiv\sum_{n=1}^\infty {\mathscr{J}}_n,\ \ \ \ \ {\mathscr{J}}_n\equiv\sum_\phi\frac{\partial {\mathscr{L}}_n}{\partial \partial_n\phi}{\mathbf{U}}_n\phi,
\end{equation}
in our notation is \cite{KO}
\begin{equation}
{\mathbf{Q}}_{(\ref{eq:trans})}=\sum_{n=1}^\infty\Biggl[\sum_{I=1}^3\biggl\{{\mathscr{G}}({\mathscr{A}}\times_G{\mathscr{C}})^I\Phi^I+i{\mathscr{G}}\partial_n\bigl({\bar{{\mathscr{C}}}}^I\bigr)\frac{{\mathscr{R}}^I}{2}\biggr\}-(\partial_n{\Phi}){\mathscr{C}}\Biggr]\;.
\end{equation}
where the contribution from ${\mathscr{A}}$ vanishes due to its equation of motion, and we rewrite the one from $\bar{{\mathscr{C}}}$ as the final term using the Leibniz rule for $\partial_n(\Phi{\mathscr{C}})$.

We have introduced the quadratic operator ${{{{\mathscr{R}}}}}$ defined by the $G=SL(2,{\mathbb{R}})$ Lie bracket
\begin{equation}{{{{\mathscr{R}}}}}\equiv {{\mathscr{C}}}\times_G{{\mathscr{C}}}\end{equation}

Due to the equation of motion of $\Phi$ and the fact that $q_n$ and $\bar{q}_n$ are nilpotent, we have
\begin{eqnarray}
 \Phi=i\sum_{n=1}^\infty\partial_n{\mathscr{A}}={{Q}}{\mathscr{A}},\ \ \ \ \ {{Q}}\equiv i\sum_{n=1}^\infty\bigl(q_n+\bar{q}_n\bigr).
\end{eqnarray}

The last term of ${\mathbf{Q}}_{(\ref{eq:trans})}$ becomes \begin{eqnarray}-\sum_{n=1}^\infty\Bigl(\partial_n\Phi\Bigr){\mathscr{C}}=-\Bigl(\Delta\Bigl[\vec{t},\vec{\bar{t}}\Bigr]{\mathscr{A}}\Bigr){\mathscr{C}}=\Lambda{\mathscr{A}}{\mathscr{C}}\end{eqnarray}

The canonical momenta of ghosts and anti-ghost are 
\begin{eqnarray}
\Pi_{{\mathscr{C}}}^I= \sum_{n=1}^\infty D_n{\mathscr{C}}^I\equiv({\partial}_{{t}}{\mathscr{C}})^I+i{\mathscr{G}}({\mathrm{curl}}_{{\varepsilon}}{\mathscr{C}})^I
\end{eqnarray}
and
\begin{eqnarray}
\Pi_{\bar{{\mathscr{C}}}}^I=i{\mathscr{G}}\sum_{n=1}^\infty{\partial_n{\bar{{\mathscr{C}}}}}^I\equiv i{\mathscr{G}}({\partial}_t\bar{{\mathscr{C}}})^I.
\end{eqnarray}
The first term of $\Pi_{{\mathscr{C}}}$ vanishes if the gauge $SL(2,{\mathbb{R}})$ symmetry is not spontaneously broken for our coherent solutions.

Using these new variables, the BRST charge is rewritten as 
\begin{equation}{\mathbf{Q}}=\sum_{I=1}^3\biggl(({\mathrm{curl}}_{\varepsilon}{\mathscr{C}})^I{\mathscr{G}}{{Q}}{\mathscr{A}}^I+({\partial}_{t}{\bar{\mathscr{C}}})^I{\mathscr{G}}\frac{i}{2}{{\mathscr{R}}}^I-\Lambda({\nabla}_{\varepsilon}{\mathscr{C}})^I\biggr)\end{equation}

We pay attention to the fact that ${\mathscr{A}}^I$ is just an operator and we will express the BRST constraint ${\mathbf{Q}}\psi=0$ on the Hilbert space
\begin{equation}{\mathscr{H}}\equiv\bigoplus_{I=1}^3\bigoplus_{J=1}^3 \Bigl<{{\nabla}_{{\varepsilon}^I}{\mathscr{C}}^J}\Bigr>.\end{equation}

We define two sets of three $3\times 3$ matrices as the elements of ${\mathscr{H}}$
\begin{equation}\Theta_{{\mathrm{flat}}}^1\equiv\left(\begin{array}{ccc}1&0&0\\0&0&0\\0&0&0\end{array}\right),\ \ \ \ \ \Theta_{{\mathrm{flat}}}^2\equiv\left(\begin{array}{ccc}0&0&0\\0&1&0\\0&0&0\end{array}\right),\ \ \ \ \ \Theta_{{\mathrm{flat}}}^3\equiv\left(\begin{array}{ccc}0&0&0\\0&0&0\\0&0&1\end{array}\right),
\end{equation}
\begin{equation}\Theta_{{\mathrm{curv}}}^1\equiv\left(\begin{array}{ccc}0&0&0\\0&0&-1\\0&1&0\end{array}\right),\ \ \ \ \ \Theta_{{\mathrm{curv}}}^2\equiv\left(\begin{array}{ccc}0&0&1\\0&0&0\\-1&0&0\end{array}\right),\ \ \ \ \ \Theta_{{\mathrm{curv}}}^3\equiv\left(\begin{array}{ccc}0&1&0\\-1&0&0\\0&0&0\end{array}\right).\end{equation}
Then, by the interchange of ghost and anti-ghost
\begin{equation}{\mathscr{C}}\leftrightarrows{\bar{\mathscr{C}}},\ \ \ \ \ \Pi_{{\mathscr{C}}}\leftrightarrows \Pi_{\bar{\mathscr{C}}}\end{equation}
the ${\mathscr{C}}$ part of the BRST charge before this interchange is equal to the $\bar{{\mathscr{C}}}$ part of the BRST charge after this interchange. That is
\begin{equation}\Theta^I_{{\mathrm{curv}}}{\mathbf{U}}{\mathscr{C}}=\Theta_{{\mathrm{flat}}}^I{\mathbf{U}}{\bar{\mathscr{C}}}\end{equation}

So, due to this symmetry on interchange, the sum of ghost and anti-ghost parts of the BRST charge is written as
\begin{equation}{\mathbf{Q}}^I=\Theta^I{\mathbf{U}}\bigl({{\mathscr{C}}}+{\bar{{\mathscr{C}}}}\bigr)^I,\ \ \ \ \ \Theta^I\equiv\frac{\Theta_{{\mathrm{flat}}}^I+\Theta_{{\mathrm{curv}}}^I}{2}\end{equation}

Multiplying by the $SL(2,{\mathbb{R}})$ coupling constant $\alpha^\prime$ we find the equation for BRST quantization is
\begin{equation}{\Biggl(\sum_{I=1}^3\Theta^I\biggl({{Q}}{\mathscr{A}}^I+\frac{i}{2}{\mathscr{R}}^I\biggr)-\alpha^\prime \Lambda\Biggr)\psi\Bigl[g_s,\vec{t},\vec{\bar{t}}\Bigr]=0}\label{eq:brst}\end{equation}
Obviously, the solution $\psi$ is 3-vector but we suppress the index of this vector in Equation~(\ref{eq:brst}).

Based on the formulation so far, we propose a conjecture about the equivalence between our model and matrix theory.

 As shown in the paper of Ishibashi et al,\cite{IKKT} the BFSS matrix model is included in the IIB matrix model. So, we will deal with the IIB matrix model, as defined by an equivalence of its path-integral and the path-integral of Schild-gauge type IIB string theory, using a large $N$ matrix model reduction of the degrees of freedom.\cite{IKKT,EK}

The bosonic part of the action of the size $N$ IIB matrix model with bosonic variables ${\mathscr{X}}_{\mu}^N$ is \begin{equation}S[g,{\mathscr{X}}_\mu]=
-\frac{1}{g^2}\biggl(\frac{1}{4}{\mathrm{Tr}}[{\mathscr{X}}_\mu,{\mathscr{X}}_\nu]^2\biggr)
\end{equation} 
The partition function for $g\equiv g_s\ell$ is defined by
\begin{equation}
{\mathscr{Z}}[g]\equiv\sum_{N=1}^\infty z_N[g],\ \ \ \ \ 
z_N[g]\equiv\int \prod_{\mu}d {\mathscr{X}}_{\mu}^{N}e^{-S[g,{\mathscr{X}}_{\mu}^{N}]}
\end{equation}

We decompose the space of size $N$ Hermitian matrices, $Her$, into sets of its irreducible sub matrices, $Her_{irr}$. The deformation of the way of decompositions is defined by
\begin{equation}\widetilde{{{Her}}}(N)=\bigcup_{I=1}^N {\mathbb{C}}_I\times\Biggl(\bigcup^{N!/((N-I+1)!I!)}_1 {{Her}}_{irr}(I)\Biggr),\ \ \ \ \ {{Her}}_{irr}(1)={\mathbb{R}}\end{equation}
We define the partition function $z_N[g]$ on ${{Her}}_{irr}(N)$ with deformation parameter $t_I\in {\mathbb{C}}_I$.

Once we choose the way of the deformations $\widetilde{Her}$ by the parameters $\vec{t}$ and $\vec{\bar{t}}$, the partition function varies with these parameters.

Then we can write the time-dependent IIB matrix model partition function as ${\mathscr{Z}}\Bigl[g,\vec{t},\vec{\bar{t}}\Bigr]$.

We conjecture that, for certain choices of gauges of $U(N)$ gauge symmetries, this partition function ${\mathscr{Z}}\Bigl[g,\vec{t},\vec{\bar{t}}\Bigr]$ is a solution of our $SL(2,{\mathbb{R}})$ BRST equation
\begin{equation}
{\mathbf{Q}}{\mathscr{Z}}\Bigl[g,\vec{t},\vec{\bar{t}}\Bigr]=0.
\end{equation}

Once the equivalence to matrix theory is established, how do we describe gravity?

Unfortunately, there is no variable for the graviton in the $S$-duality multiplet. So, we use $AdS/CFT$ duality to produce the graviton sector.

In Equation~(\ref{eq:brst}), {{gravity}} is described as follows. There are critical points $g^I$ for $I=1,2,\ldots$ with vacuum expectation values (VEV) found by substituting the saddle point value \begin{equation}|g^1|<|g^2|<\ldots\end{equation} into the gauge potential \begin{equation}{{\mathscr{Z}}^C}={\mathscr{Z}}_{\vec{m},\vec{n}},\ \ \ \ \ \ \ {\mathscr{Z}}\Bigl[g,\vec{t},\vec{\bar{t}}\Bigr]=\sum_{\vec{m}}\sum_{\vec{n}}\pm {\mathscr{Z}}_{\vec{m},\vec{n}}\Bigr[g,\vec{t},\vec{\bar{t}}\Bigr]\prod_{\vec{m}}\theta_m\prod_{\vec{n}}\bar{\theta}_n.\end{equation} That is, the component fields of the partition function ${\mathscr{Z}}$ in the superspace description satisfy
\begin{equation}
\Bigl<{{\mathscr{Z}}^C}\Bigl[{\mathbf{U}}^\perp g^N,\vec{t},\vec{\bar{t}}\Bigr]\Bigr>_N=0.
\end{equation}
for the transformation ${\mathbf{U}}^\perp$ which is orthogonal to the BRST transformation ${\mathbf{U}}$:
\begin{equation}
[{\mathbf{U}},{\mathbf{U}}^\perp]=0,\ \ \ \ \ \Bigl<1,{\mathbf{U}},{\mathbf{U}}^\perp\Bigr>=SL(2,{\mathbb{R}}).
\end{equation}

The partition function factorizes into a symmetric part ${\mathscr{Z}}_\Gamma$ and the remainder
\begin{equation}
{\mathscr{Z}}=\Bigl(\Bigl<{\mathscr{Z}}\Bigr>+\widetilde{{\mathscr{Z}}}\Bigr)\otimes {\mathscr{Z}}_\Gamma
\end{equation}

The Goldstone mode associated with the $N$-th vacuum is defined by the difference between the gauge potential and its VEV. The $SL(2,{\mathbb{R}})$ transformations of these modes
\begin{equation}
\widetilde{{\mathscr{Z}}_N}\Bigl[{\mathbf{U}} g^N,\vec{t},\vec{\bar{t}}\Bigr]\equiv
\sum_{\vec{m}}\sum_{\vec{n}}{\mathbf{U}}\widetilde{{{\mathscr{Z}}^C_N}}_{\vec{m},\vec{n}}\Bigl[g^N,\vec{t},\vec{\bar{t}}\Bigr]\prod_{\vec{m}}\theta_m\prod_{\vec{n}}\bar{\theta}_n.
\end{equation}
are defined on $\mathfrak{h}$.

For a certain neighborhood $U_N$ around $g_s$, the Goldstone mode satisfies our BRST equation locally
\begin{equation}{\mathbf{Q}}\widetilde{{\mathscr{Z}}_N}\Bigr|_{U_N}\Bigl[g,\vec{t},\vec{\bar{t}}\Bigr]=0.\end{equation}

Based on the discussion so far, we now show how to consider the large-$N$ dualities in the DGSSB framework. 

In the limits of large $N$ and 'tHooft strong coupling $g\to\infty$, the $D$-particle becomes light and, at the {minimum} of the potential can {{condense}} at any space-time point. Thus in the plane wave limit, $D$-particle field theory is able to describe the graviton as closed string fields trapped at their boundaries on $D$-particles.
 
So, the graviton is, in our sense, a Goldstone mode of the partition function $\widetilde{{\mathscr{Z}}^C_N}$ around its critical coupling $g^N$, i.e., the minimum of the gauge potential.

To clarify this issue, we pay attention to the assumption about the coupling constant ${\mathscr{G}}$ for $D$-particle fields given by Eq. (\ref{eq:G}).

In the classical gravity limit of $\alpha^\prime\to0$, the $D$-particle fields become strong coupling. Then every distance $r$ between $D$-particles is proportional to
\begin{equation}r\stackrel{\alpha^\prime\to 0}{\propto} {\mathscr{G}}^{-1}=\alpha^\prime
\end{equation}
This constrains the mass of open strings attached to $D$-particles to be constant.

Then, in the large $N$ limit (where ${g}\simeq1/g=\lim_{N\to\infty}1/\bigl(k/N\bigr) = \infty$ fixing 'tHooft's coupling $k=gN$), $AdS/CFT$ correspondence implies that\cite{M}
\begin{equation}
\lim_{N\to\infty}\widetilde{{{\mathscr{Z}}_N}}\Bigl[g,\vec{t},\vec{\bar{t}}\Bigr]
\end{equation}is the partition function of gravitons.

\smallskip
\smallskip

\leftline{{\textbf{Acknowledgment}}}

\smallskip

I wish to dedicate this article to my collaborator Jnanadeva Maharana and thank him for his kind instruction in our forthcoming work on GSSB.

\end{document}